# Indirect Quantum Approximate Optimization Algorithms: application to the TSP

Eric Bourreau[a], Gérard Fleury[b] and Philippe Lacomme[b]

[a] Université de Montpellier, LIRMM, 161 Rue Ada, 34095 Montpellier, France - eric.bourreau@lirmm.fr
[b] Université Clermont Auvergne, Clermont Auvergne INP, UMR 6158 LIMOS, 1 rue de la Chebarde, Aubière, 63178, France – gerard.fleury@isima.fr and philippe.lacomme@isima.fr

**Abstract**

We propose an Indirect Quantum Approximate Optimization Algorithm (referred to as IQAOA) where the Quantum Alternating Operator Ansatz takes into consideration a general parameterized family of unitary operators to efficiently model the Hamiltonian describing the set of string vectors. This algorithm creates an efficient alternative to QAOA, where: 1) a Quantum parametrized circuit executed on a quantum machine models the set of string vectors; 2) a Classical meta-optimization loop executed on a classical machine; 3) an estimation of the average cost of each string vector computing, using a well know algorithm coming from the OR community that is problem dependent. The indirect encoding defined by dimensional string vector is mapped into a solution by an efficient coding/decoding mechanism.
The main advantage is to obtain a quantum circuit with a strongly limited number of gates that could be executed on the noisy current quantum machines. The numerical experiments achieved with IQAOA permits to solve 8-customer instances TSP using the IBM simulator which are to the best of our knowledge the largest TSP ever solved using a QAOA based approach.

## 1. Introduction

Quantum optimization stands as a promising and innovative field with the potential of significant implications in the realm of operations research. This emerging frontier allows us to tackle minimization problems using quantum metaheuristics, promising a highly effective approach that overcomes the challenge of getting stuck in local minima—a common issue faced by conventional local search algorithms.

The efficacy of methods grounded in Simulated Annealing, commonly employed in the field of operations research, hinges on the gradual reduction of a parameter '$t$' to zero, facilitating the traversal of potential energy barriers. Various metaheuristic approaches exhibit distinct strategies to prevent premature convergence to local minima while sustaining robust capabilities for efficient exploration of the search space. An outstanding representative within this diverse category of metaheuristics, which encompasses a wide array of techniques such as memetic algorithms, GRASP, and VNS, is the Simulated Annealing method. Collectively, these metaheuristic techniques augment the arsenal available for tackling optimization challenges in the ever-evolving realm of operations research.

From the perspective of quantum mechanics, quantum fluctuations bear resemblance to thermal fluctuations. What sets quantum mechanics apart from classical approaches is the capability of waves to tunnel through potential energy barriers, as expounded by Martoňák et al. in 2004 (Martoňák et al., 2004). In recent years, the quantum physics community has introduced several quantum metaheuristics, giving rise to a family of quantum approximate algorithms. Notable among these is the Adiabatic-based Algorithms, which offer an approximate solution to the Schrödinger equation, as formulated by Schrödinger in 1926 (Schrödinger, 1926).

Recently, Farhi et al. (2015) introduced a new class of algorithms centered around the alternation between two distinct sets of operators: Hamiltonian and mixing Hamiltonian. This alternation process gives rise to Quantum Approximate Optimization Algorithms, commonly referred to as QAOA. These algorithms represent a hybrid approach in which the classical computer is tasked with exploring the search space to optimize a set of parameters, while the evaluation of probability distributions is executed by a quantum device.

It's noteworthy that QAOA does not support local search considerations and offers a comprehensive exploration of the entire search space. This pioneering work was further expanded upon in the renowned publication by Hadfield in 2018 (Hadfield, 2018). In their work, they introduced new ansatz that specifically facilitate the exploration of the feasible subspace, ensuring that hard constraints are inherently satisfied. This approach bears a striking resemblance to classical methodologies within the Operations Research (OR) community, as it involves a meticulous definition of classical operations, such as qubit permutations within the qubit-string used for solution modeling.

## 2. Indirect Quantum Approximate Optimization Algorithms (IQAOA)

### 2.1. Mapping function in OR field

A great challenge in OR consists in generating consistent model that represent solutions only. Investigating only feasible subspace has received attention to the Operation Research (O.R.) community for decades and such approaches have been successfully applied in several research areas leading to very efficient metaheuristics. In VRP a well-known representation is the giant trip that can be transformed into one VRP solution using the Split Algorithm (Beasley, 1993) (Lacomme et al., 2001). The Split algorithm defines a mapping from the set of giant trip of a TSP (Traveling Salesman Problem) to solutions of the VRP and the metaheuristic based approaches manipulate efficiently the set of giant trips only.

Cheng et al. in 1996 (Cheng et al., 1996) are the very first who made a full analysis of non-string coding approach and decoding mechanisms in the global context of constraint optimization. Different mappings are represented on the Figure 1.

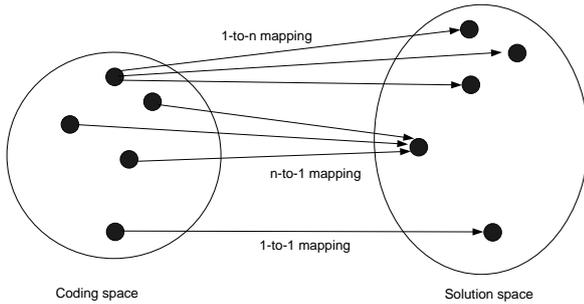

Fig. 1. Mapping from coding to solution space

In conclusion, it is noteworthy that many OR (Operations Research) problems are effectively solved through indirect representations of solutions. The principle is to explore not the entire set of solutions but the set of representations (such as the set of giant tours for VRP. The objects used for an indirect representation of solutions are typically vectors enabling the definition of mapping functions that are in $O(n)$.

The most significant advantage of these indirect representations lies in the fact that the metaheuristic explores the set of indirect representations, which is generally smaller in number than the set of solutions. This shifts the burden of problem modeling complexity away from the metaheuristic itself and onto the mapping function.

### 2.2. Indirect representation of solutions: proposition for permutations

In the field of combinatory, the Lehmer code provides a different method for encoding every conceivable permutation within a sequence comprising $n$ numbers. This code constitutes an illustrative instance of a system used for enumerating permutations and serves as a prime example of an inversion table. The term "Lehmer code" pays homage to Derrick Henry Lehmer (Lehmer, 1962), although its existence is dating back to at least 1888 (Laisant, 1888).

Multiple methods exist for establishing this one-to-one correspondence, with the most classic among them being the Lehmer code, also known as the inversion table. An algorithmic description is introduced first in (Knuth, 1981).

Indirect representation can take advantages of the one-to-one correspondence between permutations and the so-called subexceedant functions and by consequences with one integer number modeling the rank of the permutation.

Assume $f$ is a bijection that associate each permutation over the interval $[n] = \{0,1, \dots n-1\}$:
$$f: [n] \to [n]$$

$f$ is the subexceedant (Laisant, 1888) (Mantaci and Rakotondrajao, 2001) function defined by:

$f(i)$ : is the number of indices $j < i$ such that $\sigma_j < \sigma_i$

Obviously the following remarks holds (subexceedant function):
$$\forall i = 1..n, 0 \leq f(i) \leq i$$

This property justifies the term 'subsequent function' a term that can be found, for example, in (Dumont and Viennot, 1980).

Moreover we have: $f(0) = 0$

Let us note $f$ denoted by:
$$f \sim [f(n-1); f(n-2); \dots; f(1); f(0)]$$
Let us note $F_n$ the set of functions satisfying the previous condition: $Card(F_n) = n!$

**Example**
$$F_2 = \{00, 10\}$$
$$F_3 = \{000, 100, 200, 010, 110, 210\}$$

Let us consider a permutation $\sigma$ in $f$. The subexceedant function $f$ related to $\sigma$ can be obtained by iteratively scanning $\sigma$ and by assigning $f[i] = \sigma[i]$ at each iteration. The remaining elements of $\sigma$ that occurs on the right of $i$, such that $\sigma[j] > f(i)$, have to be decreased of one unit, to ensure that at the position $i+1$ to $n$ the number are in the interval $[0; n-i]$.

```
Algorithm 1. Compute_f ()
Input parameters:
    σ : a permutation of n elements
    [n] : the interval
Output parameters:
    f : the subexceedant function
Begin
    For i = n − 1 to 1 do
        f[i] = σ[i]
        For j = i − 1 to 0 do
            If (σ[j] > i) then
                σ[j]= σ[j] − 1
            Endif
        EndFor
    EndFor
    Return f
End
```
Algorithm 1. Conversion of $\sigma$ into a subexceedant function $f$

**Example**
Let us consider $\sigma = [2; 1; 0; 3]$
At iteration 1 the number $\sigma[1] = 2$ is added to $f$
$$\sigma = [2; 1; 0; 2]$$
$$f = [2, \_, \_, \_]$$
Since no remaining number of $\sigma$ is larger than 2, $\sigma[2], \sigma[3]$ are not updated but $\sigma[4]$ is decrease of one unit.
$$\sigma = [\_; 1; 0; 2]$$
$$f = [2, \_, \_, \_]$$
.
At iteration 2 the number $\sigma[2] = 1$ is added to $f$ and the two last digits 0 and 1 are iteratively investigated: 2 which is greater to 1 is decreased of one unit.
$$\sigma = [\_; 1; 0; 2]$$
$$f = [2, \_, \_, \_]$$
hence
$$\sigma = [\_; \_; 0; 1]$$
$$f = [2, 1, \_, \_]$$

At iteration 3 the number $\sigma[3] = 0$ is added to $f$ and the last digits 1 is decreased of one unit.
$$\sigma = [\_; \_; 0; 1]$$
$$f = [2, 1, 0, \_]$$
hence
$$\sigma = [\_; \_; \_; 0]$$
$$f = [2, 1, 0, \_]$$

At iteration 4 the number $\sigma[4] = 0$ is added to $f$.
$$\sigma = [\_; \_; \_; \_]$$
$$f = [2, 1, 0, 0]$$

To conclude, $f = [2, 1, 0, 0] = [f(3), f(2), f(1), f(0)]$ is the subexceedant function associated to $\sigma = [2; 1; 0; 3]$.
□

Conversely, given a subexceedant function $f$, it is possible to calculate the associated permutation using algorithm 2.

```
Algorithm 2. Compute_Permutation()
Input parameters:
    f : a subexceedant function
    [n] : an interval
Output parameters:
    σ : a permutation of n elements
Local parameters:
    v : an ordered list of n elements beginning at 0
Begin
    v = [n − 1, n − 2, … , 1,0]
    σ = []
    For i = n − 1 to 0 do
        x = f(i)
        y = v(x)
        σ[i] = y
        v = v − {y}
    EndFor
    Return σ
End
```
Algorithm 2. Computation of $\sigma_f$

**Example**
Let us $\sigma = [5,1,4,0,2,3]$ and $f = [\_, \_, \_, \_, \_, \_]$

At iteration 1 the number $\sigma[1] = 5$ is added to $f$
$\sigma = [\_, 1, 4, 0, 2, 3]$ and $f = [5, \_, \_, \_, \_, \_]$
At iteration 2 the number $\sigma[2] = 1$ is added to $f$
$\sigma = [\_, \_, 3, 0, 1, 2]$ and $f = [5, 1, \_, \_, \_, \_]$
At iteration 3 the number $\sigma[3] = 3$ is added to $f$
$\sigma = [\_, \_, \_, 0, 1, 2]$ and $f = [5, 1, 3, \_, \_, \_]$
At iteration 4 the number $\sigma[4] = 0$ is added to $f$
$\sigma = [\_, \_, \_, \_, 0, 1]$ and $f = [5, 1, 3, 0, \_, \_]$
At iteration 5 the number $\sigma[5] = 0$ is added to $f$
$\sigma = [\_, \_, \_, \_, \_, 0]$ and $f = [5, 1, 3, 0, 0, \_]$
At iteration 6 the number $\sigma[6] = 0$ is added to $f$
$\sigma = [\_, \_, \_, \_, \_, \_]$ and $f = [5, 1, 3, 0, 0, 0]$

To conclude, $f = [5, 1, 3, 0, 0, 0]$ is the subexceedant function associated to $\sigma = [5, 1, 4, 0, 2, 3]$.
□

*2.3. Indirect representation of solutions: bijection with permutation over interval $[n]$*

Establishment of one-to-one correspondence with a permutation over the interval $[n]$ and a function $f: [n] \to \{0, \ldots, n-1\}$

**Property**
For any $x \in \mathbb{N}$, we have the decomposition
$$x = \sum_{i=0}^{n-1} x_i \cdot (i!)$$
with $0 \leq x_i < i!$ to ensure unicity.
Hence $f = (x_{n-1}, x_{n-2} \ldots x_0)$ is subexceedant
□

**Example**
For example, we have:
$208 = 1 \times 5! + 3 \times 4! + 2 \times 3! + 2 \times 2! + 0 \times 1! + 0 \times 0!$
And
$$f = (1; 3; 2; 2; 0; 0) \text{ is subexceedant}$$

For example, we have:
$10 = 1 \times 3! + 2 \times 2! + 0 \times 1! + 0 \times 0!$
And
$$f = (1; 2; 0; 0) \text{ is subexceedant}$$

For a set of $n$ customers, we have $n!$ permutations numbered from 0 to $n! - 1$.
Let us consider $x \in [0; n! - 1]$ that models a rank in the list of permutations. To any rank $x \in [0; n! - 1]$, it is possible to defined the subexceedant $f$ composed of decomposition in the factorial basis and by consequence the permutation $\sigma$. To conclude for any any rank $x \in [0; n! - 1]$ (Figure 2), we can compute:
- the subexceedant function by decomposing $x$ in the factorial base ;
- the permutation $\sigma_{f(x)}$ associated to the subexceedant function $f(x)$ (algorithm 2) ;
- the cost of any permutation $\sigma_{f(x)}$ considering: $cost = \sum_{i=0}^{n-2} d_{\sigma_i, \sigma_{i+1}} + d_{\sigma_{n-1}, \sigma_0}$ assuming $d_{i,j}$ is the distance from customer $i$ to customer $j$.

For convenience we denote:
$m(x)$ the cost of permutation $\sigma_{f(x)}$ that is related to the rank $x$.
$\sigma_x$ as the permutation associated with rank $x$ whereas the correct notation should be $\sigma_{f(x)}$.

This allows us to define the mapping function that associates a permutation with each rank $x$. Cheng et al. (1996) pointed out that the most interesting mapping functions are the one-to-one functions, as they correspond to bijections between the two sets (indirect encoding and the solutions). Note that the mapping function just defined is indeed a one-to-one function, unlike the functions commonly used in Operations Research, which are of the $n$-to-one type (including for example the Split method in the VRP that is clearly of the $n$-to-one type).

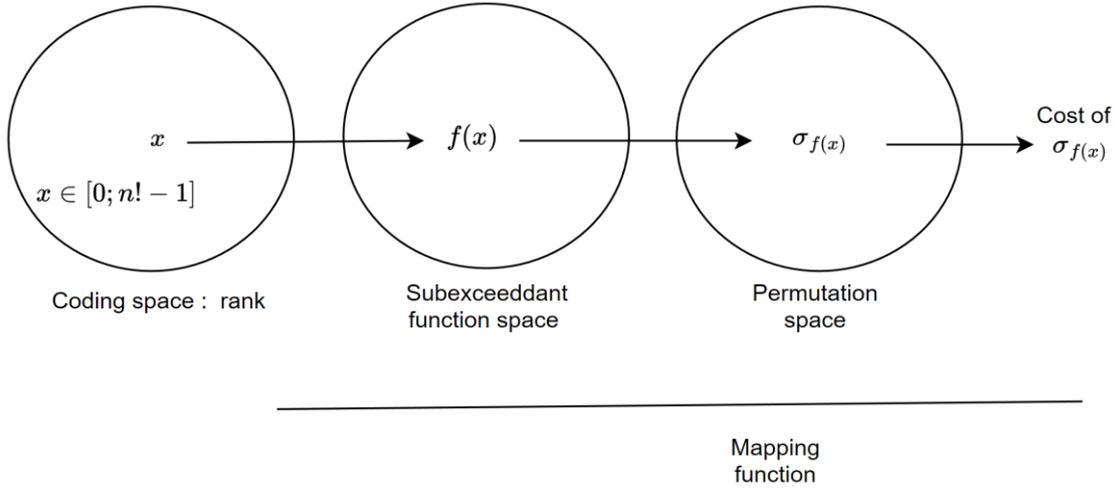

Fig. 2. Mapping from coding to solution space

By consequence, the cardinality of the code space (number of ranks) is equal to the number of permutations.

For example, for $n = 4$, we have $4! = 24$ permutations numbered from 0 to 23. The full list of permutations is provided on table 1. The rank 10 is associated to the subexceedant function $f = (1; 2; 0; 0)$ since $10 = 1 \times 3! + 2 \times 2! + 0 \times 1! + 0 \times 0!$, and the subexceedant function $f$ is associated to $\sigma = [1; 3; 0; 2]$.

The cost associated with $\sigma$ is $d_{1,3} + d_{3,0} + d_{0,2} + d_{2,1}$ i.e. the sum of the distance from customer 1 to 3, plus the distance from customer 3 to 0, plus the distance from customer 0 to 2 plus distance from customer 2 to 1.

Table 1. Full list of permutations for $n = 4$

| Rank | Permutation | Rank | Permutation |
|---|---|---|---|
| 0 | 0 1 2 3 | 12 | 2 1 0 3 |
| 1 | 0 1 3 2 | 13 | … |
| 2 | 0 2 1 3 | 14 | … |
| 3 | 0 2 3 1 | 15 | 2 1 3 0 |
| 4 | 0 3 1 2 | 16 | … |
| 5 | 0 3 2 1 | 17 | … |
| 6 | 1 0 2 3 | 18 | … |
| 7 | … | 19 | … |
| 8 | … | 20 | … |
| 9 | … | 21 | 3 1 2 0 |
| 10 | 1 3 0 2 | 22 | … |
| 11 | … | 23 | … |

### 2.4. IQAOA based approach

Quantum Approximate Optimization Algorithms (Farhi E. and Goldstone, 2014) take advantage of alternations between the cost function investigation which is modeled by a Hamiltonian $H_P$ from one side and a driver Hamiltonian operator $H_D$. The Quantum Alternating Operator Ansatz (Hadfield, 2018) takes into consideration a general parameterized family of unitary operators and create an efficient alternative to the Adiabatic Optimization.

As introduced by (Schrödinger, 1926) the wave function evolution of a quantum-mechanical system is given by

$$\frac{\partial}{\partial t}|\psi(x,t)\rangle = -\frac{i}{\hbar}.H(t).|\psi(x,t)\rangle$$

where the energy is defined by $H(t)$, $\hbar$ is derived from Plank constant and $|\psi(x,t)\rangle$ are states vectors. If $H$ is time independent the solution is $|\psi_t\rangle = e^{-\frac{i}{\hbar}t.H}.|\psi_0\rangle$. Note that the solution is $|\psi_T\rangle = e^{-\frac{i}{\hbar}\int_o^T H(u).du}.|\psi_0\rangle$ in the general time dependent situation. Describing a problem with a Hamiltonian $H$ and an initial state $|\psi_0\rangle$ allows to compute the ground state. The time-dependent Schrödinger's equation can be explicitly solved in very specific situation (see (Kadowaki T. and Nishimori, 1998), for one example).

### 2.5. Modelling rank and search space investigation

IQAOA seeks to solve a hard optimization problem i.e. minimizing or maximizing one objective function $m(x)$ that is assumed to act on $n - bits$ strings that model only the rank of one solution. IQAOA is based on $p$ consecutive iterations of one Hamiltonian $H_P$ cumulated with a driver Hamiltonian $H_D$, where this weighted sum of Hamiltonian terms varies in time. The Hamiltonian $H$ maps the function the rank $x$ with $2^n$ eigenvalues that model the $2^n$ values of the rank.

The Hamiltonian is implemented into a quantum circuit by deriving $U_H(t) = e^{-i.H.t}$ with $t \in [0; 2\pi]$ and using only and Z-rotations and $t$ refers to the weight in the iterative search process of QAOA. This permits to model all the rank of the TSP.

### 2.6. Search space investigation

The $\vec{\beta}$ and $\vec{\gamma}$ weights parametrized a quantum state $|\varphi(\vec{\beta},\vec{\gamma})\rangle$ that defines a solution rank $x$ with probability $|\langle x||\varphi(\vec{\beta},\vec{\gamma})\rangle|^2$

and an expectation value $\langle \varphi(\vec{\beta},\vec{\gamma})|C^p|\varphi(\vec{\beta},\vec{\gamma})\rangle$ estimated by sampling. Each sampling gives a measure that is a rank in the list of TSP solution that can be evaluated using the $1-to-1$ function into the associated subexceedant function first, into the permutation $\sigma_{f(x)}$ second and next the $m(x)$ cost of the permutation. This sampling permit to estimate the average cost of the problem $P$: $C^p\left(\vec{\beta},\vec{\gamma}\right)$ taking advantage of the mapping function.

The quantum computer is used to construct the state:

$$|\varphi(\vec{\beta},\vec{\gamma})\rangle = e^{-i.\vec{\beta}.H_D.t}.e^{-i.\vec{\gamma}.H_P.t}$$

For a fixed $\vec{\beta},\vec{\gamma}$, the quantum computer is used to make the stage $|\varphi(\vec{\beta},\vec{\gamma})\rangle$ and the measure in the computational basis is achieved to get a string $x$ and evaluated $\langle \varphi(\vec{\beta},\vec{\gamma})|C^p|\varphi(\vec{\beta},\vec{\gamma})\rangle$.

The overall algorithm description is illustrated in Figure 3.

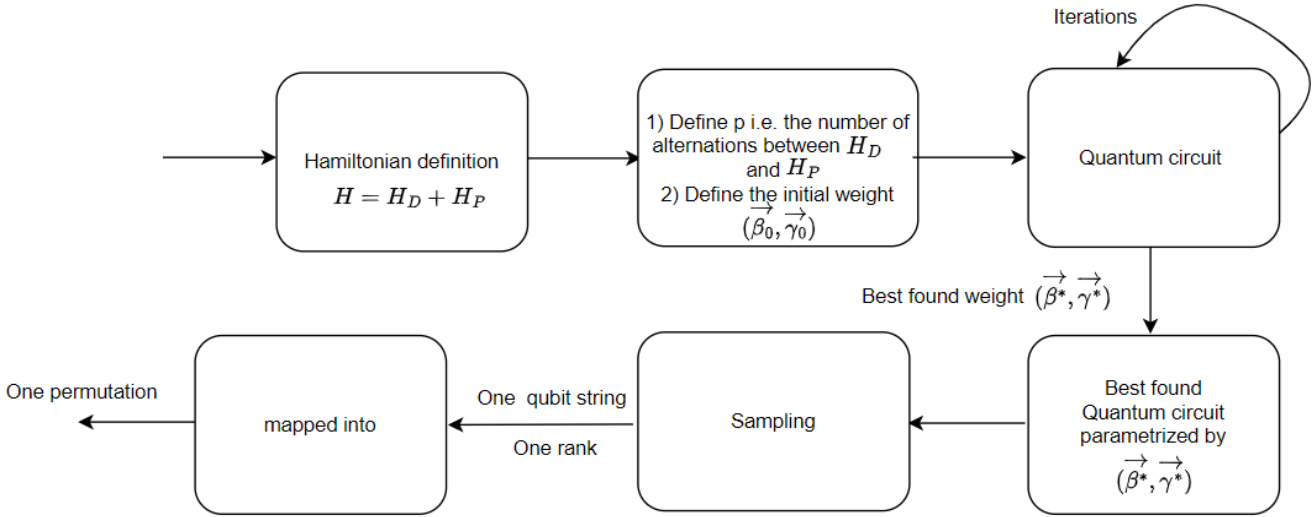

Fig. 3. IQAOA principles

IQAOA efficiency strongly relies on some key-points:
- The capacity to provide a good ratio between the estimation of quality of $C^p\left(\vec{\beta},\vec{\gamma}\right)$ versus the number of shots is required and must be tuned carefully.
- the last computed distribution $|\psi(\vec{\beta^*},\vec{\gamma^*})\rangle$ must be collected on a subset of solutions strongly smaller than the total number of solutions to avoid a costly inefficient enumerations i.e. the algorithm has converged to the optimal solutions and quasi solutions.
- The availability of one dedicated methods to computed the $(\vec{\beta^*},\vec{\gamma^*})$.
- The number of qubits $p$ such that $2^p \geq n!$ and the number of gates is very small as regards the number of qubits and gates required to encode a permutation on the quantum circuit using the classical QAOA approach.

Let us note that the expectation value $\langle \varphi(\vec{\beta},\vec{\gamma})|C^p|\varphi(\vec{\beta},\vec{\gamma})\rangle$ estimated by sampling that is the common criteria with QAOA can be replaced by more specific criteria including median, quartile or any convenient combination of this criteria depending on the objective we expect on the final distribution of probabilities.

2.7. C_GRASP×ELS for $(\vec{\beta^*},\vec{\gamma^*})$ computation

The GRASP×ELS is a fusion of two powerful algorithms: GRASP (Greedy Randomized Adaptive Search Procedure) (Feo et al., 1995) and ELS (Evolutionary Local Search) (Wolf et al., 2007; Prins, 2004). This combination joins the strengths of both methods.

The multi-start strategy of GRASP relies on a greedy randomized heuristic that generates the set of initial solutions ($np$ solutions). These solutions are then refined through a local search procedure

The second component is ELS, an extension of ILS (Iterated Local Search) introduced by (Lourenço et al. 2003). In each iteration ($ne$), a duplicate of the current solution is created, and this copy generates $nd$ child solutions. The best-performing solution among these offspring becomes the new current solution. The overarching goal of GRASP is to enhance diversity during the exploration of the global solution space, while ELS's purpose is to intensify the search within the vicinity of the current local optimum. Let us note that the method lies on local search and not on gradient and that suppose an ad hoc neighborhood definition.

**Remark 1**.
An efficient implementation of C_GRASP×ELS for Continuous GRASP×ELS required a neighboring system that consists in a proper definition of $\Delta\vec{\beta}$ and $\Delta\vec{\gamma}$.
$$(\vec{\beta},\vec{\gamma}) \rightarrow (\vec{\beta}+\Delta\vec{\beta}; \vec{\gamma}+\Delta\vec{\gamma})$$

**Remark 2**.
For efficiency reason C_GRASP×ELS can be used to optimize simultaneously $\vec{\beta}$ and $\vec{\gamma}$ first and to minimize second $\vec{\gamma}$ only.

**Remark 3**.
Depending on the desired probability distribution, it is necessary to choose the right criteria or criteria to minimize.

Among all these criteria, we can mention, without aiming to be exhaustive:
- Minimization of the expectation value of the distribution that provides insight into the central tendency;
- Minimization of the decile i.e. minimization of the data set part that contain 10% of the data;
- Minimization of the expectation value of the decile i.e. expectation value of the data that contain 10% of the data;
- Minimization of the quartile i.e. minimization of the data set part that contain 25% of the data.
- Minimization of the expectation value of the quartile i.e. expectation value of the data that contain 25% of the data;

The quartile and the decile are used to summarize the central tendency and spread of a dataset. Both quartile and decile provide a way to assess the distribution and variability of data while also identifying potential outliers and extreme values.

In the quest for a probability distribution that effectively increases probabilities around cost-efficient solutions while simultaneously mitigating the residual probabilities across a multitude of values, it is logical to explore combinations of statistical metrics, specifically the mean, and criteria intimately tied to its trend, such as deciles or quartiles, among other possibilities.

The numerical tests conducted and presented below demonstrate that it is possible to obtain a probability distribution that concentrates on high-quality solutions close to the optimal solution and even on the optimal solution itself. These tests are performed on instances ranging from 6 to 10 clients with different types of objectives to minimize and various parameters. It is worth noting that the parameters used were determined after a brief numerical study but were not subject to a specific investigation, which would be beyond the scope of this publication. All the experiments have been achieved using Qiskit (IBM) using the simulator.

### 2.8. Resolution of a TSP with 6 customers

The total number of permutations is 720 but there is only 53 different costs and the distance between customers are introduced in table 2.

Table 2. Distances

|   | 0 | 1 | 2 | 3 | 4 | 5 |
|---|---|---|---|---|---|---|
| 0 | 0 | 31 | 2 | 23 | 14 | 50 |
| 1 | 31 | 0 | 110 | 152 | 213 | 14 |
| 2 | 2 | 110 | 0 | 21 | 221 | 23 |
| 3 | 23 | 152 | 21 | 0 | 311 | 32 |
| 4 | 14 | 213 | 221 | 311 | 0 | 41 |
| 5 | 50 | 14 | 23 | 32 | 41 | 0 |

A sampling of permutations permits to show that the high quality solutions have a very low probability (Figure 4). Note that the higher probabilities are related to very low quality solutions: some solution with a cost about 800 and 900 have a probability greater than 15%.

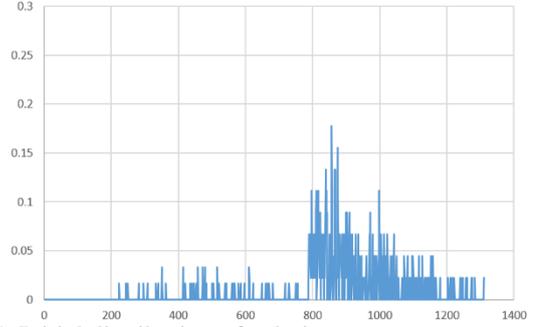
Fig. 4. Initial distribution of solutions

The instance has 12 optimal solutions (direct or reverse travel beginning by each customer) that are listed in table 3 that shows the correspondence between the rank, the permutation and the cost.

Table 3. Optimal solutions for instance with 6 customers

| Rank | Permutation |   |   |   |   |   | Cost |
|---|---|---|---|---|---|---|---|
| 55 | 1 | 4 | 3 | 2 | 6 | 5 | 223 |
| 90 | 1 | 5 | 6 | 2 | 3 | 4 | 223 |
| 150 | 2 | 3 | 4 | 1 | 5 | 6 | 223 |
| 235 | 2 | 6 | 5 | 1 | 4 | 3 | 223 |
| 286 | 3 | 2 | 6 | 5 | 1 | 4 | 223 |
| 291 | 3 | 4 | 1 | 5 | 6 | 2 | 223 |
| 376 | 4 | 1 | 5 | 6 | 2 | 3 | 223 |
| 419 | 4 | 3 | 2 | 6 | 5 | 1 | 223 |
| 494 | 5 | 1 | 4 | 3 | 2 | 6 | 223 |
| 585 | 5 | 6 | 2 | 3 | 4 | 1 | 223 |
| 632 | 6 | 2 | 3 | 4 | 1 | 5 | 223 |
| 701 | 6 | 5 | 1 | 4 | 3 | 2 | 223 |

C_GRASP×ELS is executed with the following parameters:
- Minimization of the expectation value of the decile plus the expectation value of the distribution;
- The sampling of $|\varphi(\vec{\beta},\vec{\gamma})\rangle$ is achieved with 50 shots;
- The parameters $np = 20$, $ne = 5$, $nd = 3$ for the first C_GRASP×ELS execution to optimize simultaneously $\vec{\beta}$ and $\vec{\gamma}$.
- The parameters $np = 20$, $ne = 5$, $nd = 5$ for the second C_GRASP×ELS execution to optimize $\vec{\gamma}$.
- During the local search both $\Delta\vec{\beta}$ and $\Delta\vec{\gamma}$ vary from 0.1 to 0.001 first at the beginning of the ELS. The value 0.001 is slowly decreased (divided by 10) at each iteration neighborhood generation.
- The quantum circuit is used with $p = 2$
- 40 shots are used during the optimization process to obtain a suitable evaluation of the probability distribution.
- 1000 shots are used at the end of the optimization to obtain an accurate evaluation of the probability distribution.

The landscape of the function, with ranks represented on the $x$-axis, is not a smooth landscape that facilitates the search for local minima (figure 5). However, the C_GRASP $x$ ELS method easily, with a relatively low number of iterations, finds a minimum of the function. Likewise, it seems evident that gradient-based methods will face significant challenges with this type of problem.

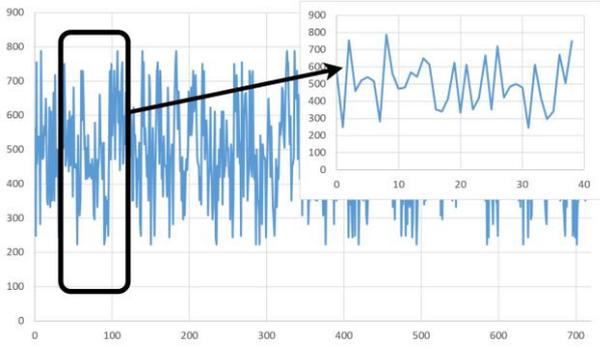

Fig. 5. Function landscape

The sampling with 1000 shots gives 1010111101 with 283 shots (figure 6) meaning that about 28% of the probabilities is now on 1010111101 that model the rank 701 and the rank number 701 is mapped into the permutation $\sigma = [6, 5, 1, 4, 3, 2]$.

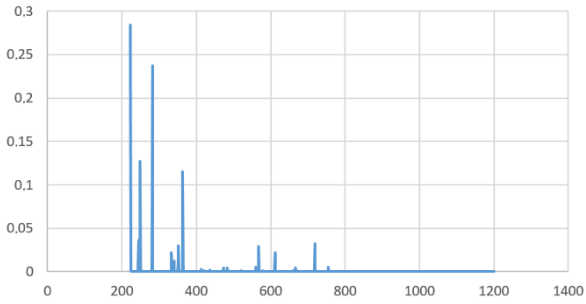

Fig. 6. Final distribution of solutions

The details provided in Table 4 confirm what the visual representation suggests, namely very high probabilities concentrated on low-cost solutions, thus demonstrating the effectiveness of the IQAOA method in solving this 6-customers TSP problem.

Table 4. Optimal solutions for instance with 6 customers

| Cost | Probability % |
|------|---------------|
| 223  | 28.4          |
| 244  | 3.6           |
| 249  | 12.7          |
| 282  | 23.7          |
| 333  | 2.2           |
| 340  | 1.2           |
| 351  | 1.4           |
| 352  | 3.0           |
| 363  | 11.5          |
| 414  | 0.3           |
| …    | …             |

*2.9. Resolution of a TSP with 8 customers*

The total number of permutations is 40320 but there are only 833 different costs and the distance between customers are introduced in table 5.

The optimal solution is 108 avec the related family permutation is:
$$\sigma = [1,2,3,4,5,6,7,8]$$

Table 5. Distances

|   | 0 | 1 | 2 | 3 | 4 | 5 | 6 | 7 |
|---|---|---|---|---|---|---|---|---|
| 0 | 0 | 10 | 21 | 23 | 40 | 115 | 66 | 17 |
| 1 | 10 | 0 | 11 | 47 | 88 | 29 | 55 | 161 |
| 2 | 21 | 11 | 0 | 12 | 22 | 123 | 24 | 25 |
| 3 | 23 | 47 | 12 | 0 | 13 | 32 | 66 | 34 |
| 4 | 40 | 88 | 22 | 13 | 0 | 14 | 42 | 33 |
| 5 | 115 | 29 | 123 | 32 | 14 | 0 | 15 | 52 |
| 6 | 66 | 55 | 24 | 66 | 42 | 15 | 0 | 16 |
| 7 | 17 | 161 | 25 | 34 | 33 | 52 | 16 | 0 |

The experiments were carried out with:
- The parameters $np = 20$, $ne = 5$, $nd = 3$ for the first C_GRASP×ELS execution to optimize simultaneously $\vec{\beta}$ and $\vec{\gamma}$.
- The parameters $np = 20$, $ne = 5$, $nd = 5$ for the second C_GRASP×ELS execution to optimize $\vec{\gamma}$.

The instance encompasses 16 optimal solutions that value 108 meaning that uniform sampling gives a probability about 0.039% to find one optimal solution (figure 7).

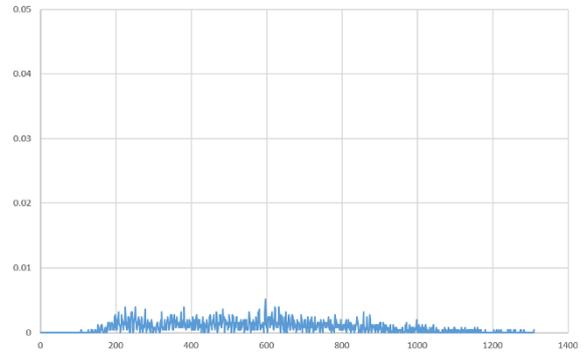

Fig. 7. Initial distribution of solutions for the instance with 8 customers.

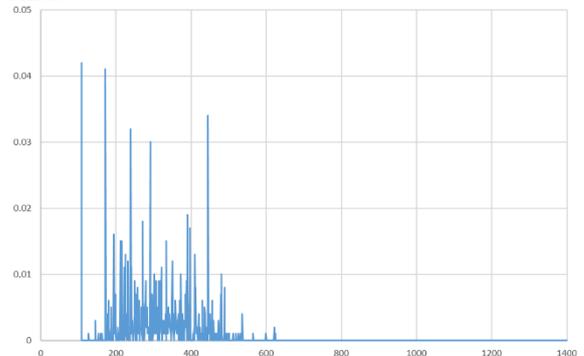

Fig. 8. Final distribution of solutions for the instance with 8 customers

The representations of the distributions in Figures 7 and 8 show that the probability distribution has been significantly altered to concentrate on high-quality solutions. It should be noted that the median is 306, which means that 50% of the data corresponds to solutions with costs lower than 306. The final sampling achieved at the end of the optimization gives probability of 4.2% associated with 108. It is challenging to provide a representation of the function to be minimized. Nevertheless, figure 9 gives a partial representation of $\vec{\beta}$ alongside the associated costs, that suggests that the function could involve numerous local minima.

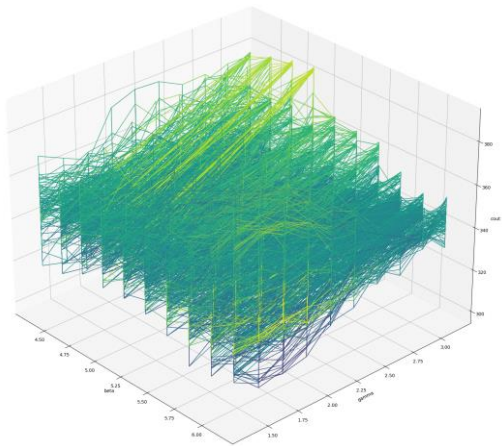

Fig. 9. Partial representation of the solution landscape just around the best solution found

## 3. Conclusion and future works

We have introduced the IQAOA approach which leverages an indirect representation of permutations. Our results offer valuable insights into the performance of IQAOA and suggest promising strategies for its practical implementation on near-term quantum devices.

To the best of our knowledge this is the first quantum resolution of a 8-customers TSP. Indirect coding including in QAOA permits to define IQAOA approach that requires a very specific method for optimization of angles that could take advantages of numerous meta-heuristic well-known in the OR-community.